\documentstyle[12pt]{article}

\begin{document}

\topmargin 0pt
\oddsidemargin 5mm
\renewcommand{\thefootnote}{\fnsymbol{footnote}}

\begin{titlepage}
\setcounter{page}{0}
\rightline{Preprint YERPHY-1412(8)-94}

\vspace{2cm}
\begin{center}
{\Large Pseudoclassical and quantum theory of the D=2n dimensional
relativistic spinning particle with anomalous ``magnetic'' moment
  in the external Yang-Mills field}
 \footnote{ Partially supported  by the grant 211-5291 YPI
 of the German Bundesministerium f\"ur Forschung und Technologie.}
 \vspace{1cm}

{\large Grigoryan G.V.,Grigoryan R.P.,Sarkissian G.A.} \\
\vspace{1cm}
{\em Yerevan Physics Institute, Republic of Armenia}\\
\end{center}

\vspace{5mm}
\centerline{{\bf{Abstract}}}
The pseudoclassical hamiltonian and action of the $D=2n$
dimensional Dirac particle with anomalous magnetic moment
interacting with the external Yang-Mills field are found. The
Bargmann-Michel-Telegdi equation of motion for the Pauli-Lubanski
vector is deduced.  The canonical quantization of $D=2n$
dimensional Dirac spinning particle with anomalous magnetic
moment in the external Yang-Mills field is carried out in
the gauge which allows to describe simultaneously particles and
 antiparticles (massive and massless) already at the classical
level.  Pseudoclassical Foldy-Wouthuysen transformation is used
to obtain canonical (Newton-Wigner) coordinates and in terms of
these variables the theory is quantized.

\vfill
\centerline{\large Yerevan Physics Institute}
\centerline{\large Yerevan 1993}

\end{titlepage}
\newpage
\renewcommand{\thefootnote}{\arabic{footnote}}
\setcounter{footnote}{0}

\section{Introduction}
\indent

As it is well known \cite{BM1,BM2},  within the pseudoclassical
approach to the theory of interaction of the point particle with
external fields Grassmann variables can be used to describe not
only spin degrees of freedom, but internal symmetries as well.
The advantage of this is that after quantization one gets
automatically finite-dimensional representations of the internal
symmetry group. Following the suggestion of \cite{BM1,BM2} in
papers \cite{BCL1,BSSW} the pseudoclassical theory of the point
particle interacting with external Yang-Mills field was
developed.

This paper is devoted to the construction of the pseudoclassical
theory of the relativistic spinning particle with anomalous
"magnetic moment" (AMM), interacting with external Yang-Mills
field and its quantization.The quantization of the relativistic
spinning particle with anomalous magnetic moment (AMM)
interacting with external electromagnetic field was investigated
in [5-8]. In  paper \cite{GG5} of the authors the
 lagrangian of the interaction
of the relativistic spinning particle with AMM with the external
electromagnetic field in $D=2n$ dimensions was constructed and
 the Bargmann-Michel-Telegdi
equation of motion for the Pauli-Lubanski vector was obtained.
The quantization of the theory was carried out in terms of
canonical (Newton-Wigner) coordinates, which were found using
pseudoclassical Foldy-Wouthuysen transformation \cite{BCLFW,GG4}.
This allowed to bypass rather cumbersome calculations of the
Dirac brackets and the subsequent diagonalization of these
brackets.

This paper is in the stream of the paper \cite{GG5} and in
essence is the generalization of it to the presence in the theory
of the internal symmetries.

In sect.2 the hamiltonian and the lagrangian (action) of the
$D=2n$-dimensional spin $\frac{1}{2}$ particle with anomalous
"magnetic moment" in the external Yang-Mills field are found and
the Bargmann -Michel -Telegdi equation for the Pauli -Lubansky
vector is deduced. In sect.3 the Foldi -Wouthuysen transformation
is used to find canonical variables, in terms of which the
quantization of the theory is carried out in sect.4.

\section{The Action of spin $\frac{1}{2}$ particle with anomalous
 magnetic moment in the external Yang-Mills field and the
 Bargmann- Michel- Telegdi equation} \indent

To describe  the internal symmetries in the  pseudoclassical
theory of the interaction  of the relativistic spinning  particle
with  the  external  Yang-Mills  field  we  introduce  as  in
[1-4]  Grassmann variables $\theta_m,\quad  m=1,\dots,N$. Let
the transformation  of the variables $\theta_m$ under the gauge
transformations of the  $G$ group have the following form:
\begin{equation}
\label{GT}
\delta \theta_m=iT^a_{mn}\theta_n \a^a(x),
\end{equation}
where $T^a$ are generators of the given representation of the
group $G$, satisfying the relations
\begin{equation}
\label{GCR}
 [T^a,T^b]=if^{abc}T^c,
\end{equation}
and which in the case of real $\theta_m$ considered here are
antisymmetric; $f^{abc}$ are the structure constants of $G$,
$\a^a(x)$  are parameters  of  the infinitesimal  gauge
transformations.   The requirement of   the invariance   of  the
theory  under   the transformations (\ref{GT}), as is well known,
necessarily leads to the introduction of the YM fields $A^a_\mu$.

 To construct the theory we postulate the constraints
\begin{eqnarray}
\label{CnstA}
&&\Phi_\mu=\pi_\mu-\frac{i}{2}\xi_\mu,\quad\mu=0,1,\ldots,D-1;\quad
\Phi_D=\pi_{D+1}+\frac{i}{2}\xi_{D+1},\nonumber\\
&&\hspace{2cm}\Phi_m=\pi_m+\frac{i}{2}\theta_m,\quad m=1,\ldots,N
\end{eqnarray}
where   $\xi_\mu,   \xi_{D+1}$   are   Grass\-mann variables
de\-scrib\-ing  the  spin  de\-grees   of  freedom ,   $\pi_\mu,$
$\pi_{D+1},$ $ \pi_m$ are the momenta, canonically conjugate to
$\xi_\mu, \xi_{D+1}$ and $\theta_m$ correspondingly. The
constraints $\Phi_\mu$ and $\Phi_D$ are the same as in the case
of the particle without AMM (free or in the external field
\cite{GG5,GG4}). The constraint $\Phi_m$ is postulated in analogy
with the constarints $\Phi_\mu,\,\Phi_D$. The Dirac brackets
${\{\ldots,\ldots\}}^*$ of the variables of the theory for the
set of constraints (\ref{CnstA}) are given by the relations
\begin{equation}
\label{PrelDB}
{\{x_\mu, P_\nu\}}^* = g_{\mu\nu},\quad
\{{\xi_\mu,\xi_\nu\}}^* = ig_{\mu\nu},\quad
 {\{\xi_{D+1},\xi_{D+1}\}}^* = -i, \quad
{\{\theta_m,\theta_n\}}^* = -i\delta^{mn},
 \end{equation}
(all other brackets vanish).  Here the variables $x^\mu$  are the
 coordinates of the particle,  $P_\mu$ are the momenta  conjugate
to $x_\mu$. The  analogue of the  fermionic constraint, which  in
the  case  of  the   relativistic  spinning  particle  with   AMM
interacting with  the external  electromagnetic field  \cite{GG5}
 after quantization using  brackets (\ref{PrelDB}) brought  to the
 covariant Dirac equation  , in this  case will be  chosen in the
form
\begin{equation}
\label{CnstD1}
\Phi_{D+1}={\cal P}_\mu\xi^\mu-m\xi_{D+1}+
iRG_{\mu\nu}^a I^a \xi^\mu\xi^\nu\xi_{D+1}
\end{equation}
where $R$ is a  parameter of the theory,($R$ - analogue of the
AMM of the particle in
electrodynamics), ${\cal  P}_\mu=P_\mu-gA_\mu^a I^a$,  $ A_\mu^a,
 G_{\mu\nu}^a$ are the vector-potential and the stress tensor  of
 the  YM   field,  $I^a$   are  the   generators  of   the  gauge
transformation .
\begin{equation}
\label{GTr}
\delta\theta=\{\theta,\a^a(x)I^a\}
\end{equation}
($\{\dots,\dots\}$ denotes the Poisson bracket), which are given
by the expressions (see \cite{BCL1, BSSW})
\begin{equation}
\label{IG}
I^a=i\pi_m T^a_{mn}\theta_n =\frac{1}{2}\theta_m T_{mn}\theta_n
\end{equation}
and satisfy the relations
\begin{equation}
\label{IGC}
\{I^a,I^b\}=f^{abc}I^c.
\end{equation}
Note, that for $T^a_{mn}$, belonging to a certain representation
(of the dimension $N$) of the $L_G$ algebra of the gauge group
$G$,  the generators $I^a$ represent a subalgebra of the algebra
$L_{SO(N)}$ of the orthogonal group $SO(N)$, and hence the group
$G$ is a subgroup of the latter \cite{BSSW}.

Considering now , as in \cite{GG5}, a theory with Dirac brackets
(\ref{PrelDB}) and a hamiltonian
\begin{equation}
\label{PrelHam}
H = \frac{i\chi}{2}\Phi_{D+1},
\end{equation}
where  $\chi$  is  the  Grassmann  odd  Lagrange  multiplier   to
 $\Phi_{D+1}$,  we  find  from  the  consistency condition  a new
  constraint $\Phi_{D+3} $
\begin{eqnarray}
\label{CnstD2}
&& {\{\Phi_{D+1},\Phi_{D+1}\}}^*  =
i\Phi_{D+3} = i({\cal P}_\mu {\cal P}^\mu -
igM G_{\mu\nu}^a I^a\xi^\mu\xi^\nu+ \nonumber\\
&& +4iRG_{\mu\nu}^a I^a{\cal P}^\mu\xi^\nu\xi_{D+1}+
R^2{(G_{\mu\nu}^a I^a\xi^\mu\xi^\nu)}^2 - m^2) \approx 0,
\end{eqnarray}
where $gM=(g-2Rm),\quad M$  is the total "magnetic" moment
of the particle. To deduce (\ref{CnstD2}) we had made use of the
relation
\begin{equation}
\label{CM}
{\{{\cal P}_\mu, {\cal P}_\nu\}}^*=gG_{\mu\nu}\equiv
gG^a_{\mu\nu}I^a,
\end{equation}
which can be easily checked by direct calculation , and also of
the identity
\begin{eqnarray}
\label{Iden}
& &\{G_{\mu\nu}, {\cal P}_{\lambda}\}^*\xi^\mu \xi^\nu \xi^\lambda
=\nabla ^{ab}_\lambda G^b_{\mu\nu}I^a\xi^\mu \xi^\nu \xi^\lambda\equiv 0,\\
& &\hspace{2cm} \nabla^{ab}_\lambda=
 \delta^{ab}\partial_\lambda+gf^{abc}A_\lambda^c\nonumber
\end{eqnarray}
which can be derived using (\ref{CM}) and the Jacoby identity
\begin{equation}
\label{JI}
\{ \{{\cal P}_\mu \xi^\mu,{\cal P}_\nu \xi^\nu\}^*,
{\cal P}_\lambda \xi^\lambda\}^* \equiv 0.
\end{equation}
Now taking into account the constraint $\Phi_{D+3}$ we can
construct the extended hamiltonian
\begin{equation}
\label{ExtHam}
H_{\rm ext} =
\frac{i\chi}{2}\Phi_{D+1}+\frac{e}{2} \Phi_{D+3},
\end{equation}
where $e$ is Grassmann even Lagrange multiplier to the constraint
$\Phi_{D+3}$.  The consistency equations imply no new constraints
since due to Jacoby identity
\begin{equation}
\label{JacobyId}
{\{\Phi_{D+3},\Phi_{D+1}\}}^*={\{{\{\Phi_{D+1},
\Phi_{D+1}\}}^*,\Phi_{D+1}\}}^*=0.
\end{equation}
Hence the dynamics of the theory is described by the hamiltonian
(\ref{ExtHam})

The  action  (lagrangian)  of  the relativistic spinning particle
with  AMM  in  the  external  YM  field  can be found by Legendre
transformation   using   (\ref{ExtHam})   and   the   constraints
(\ref{CnstA}).In this way we find that
\begin{eqnarray}
\label{Action}
 S&=& {\frac{1}{2}}\int d\tau\Bigl[\frac{\left(\dot x^\mu\right)^2}
{e}+em^2-i\left(\xi_\mu\dot \xi^\mu-
\xi_{D+1}\dot \xi_{D+1}\right)+i\theta_m\dot \theta_m +\nonumber\\
&& +2g\dot x^\mu A_\mu^aI^a+igMe
 G^a_{\mu\nu}I^a\xi^\mu\xi^\nu
 -4iR G^a_{\mu\nu}I^a {\dot x}^\mu \xi^\nu\xi_{D+1}-\\
&& -i\chi \bigl(\frac{\xi_\mu\dot x^\mu}{e}-m\xi_{D+1}
-iR G^a_{\mu\nu}I^a\xi^\mu\xi^\nu\xi_{D+1}\bigr)-
eR^2( G^a_{\mu\nu}I^a\xi^\mu\xi^\nu)^2\nonumber
\Bigr],
\end{eqnarray}
where the overdote denotes the differentiation with respect to
$\tau$ along the trajectory of the particle.

Following \cite{GG5} we write down the complete set of the
constraints of the theory
\begin{equation}
\label{C1}
\Phi_\mu=\pi_\mu-\frac{i}{2}\xi_\mu,\quad
\Phi_{D+1}=\pi_{D+1}+\frac{i}{2}\xi_{D+1},\quad
\Phi_m=\pi_m+\frac{i}{2}\theta_m,
\end{equation}
\begin{eqnarray}
\label{C2}
&&\Phi_{D+3}={\cal P}_\mu {\cal P}^\mu -
igMG_{\mu\nu}^a I^a\xi^\mu\xi^\nu+ \nonumber\\
&& +4iRG_{\mu\nu}^a I^a{\cal P}^\mu\xi^\nu\xi_{D+1}+
R^2{(G_{\mu\nu}^a I^a\xi^\mu\xi^\nu)}^2 - m^2
\end{eqnarray}
\begin{equation}
\label{C3}
\Phi_{D+4}= x^\prime _o,
\end{equation}
\begin{equation}
\label{C4}
\Phi_{D+5}=\pi_e,\quad\Phi_{D+6}=e-\frac{\kappa}{{\cal P}_o},
\end{equation}
\begin{equation}
\label{C5}
\Phi_{D+1}={\cal P}_\mu\xi^\mu-m\xi_{D+1}+
iRG_{\mu\nu}^a I^a \xi^\mu\xi^\nu\xi_{D+1},\quad
\Phi_{D+2}=a\xi_o+b\xi_{D+1}.
\end{equation}

Here $\pi_e$ is  the canonical momentum,  conjugate to the  $e$;
$\kappa=\pm  1,\kappa=+1$  corresponds  to  the  presence of the
particle in the theory,  while $\kappa=-1$ to that of
antiparticle;  $a$ and  $b$  are  parameters of  the  theory,
$a^2+b^2\neq0$.  When $a\neq0$, the theory has the massless
limit ($m \rightarrow 0$) \cite{GG1}.

The system (\ref{C1}-\ref{C5}) includes the constraints, which
follow from the action (\ref{Action}) using Dirac procedure of
the evaluation of new constraints, as
well as additional constraints, introduced into the theory for a
complete fixation of the gauges,  so that now all the constraints
are second class. The additional constraints are $\Phi_{D+4},
 \Phi_{D+6}, \Phi_{D+2}$. Note \cite{GG1}, that
 $x^\prime_o=x_o-\kappa\tau$ and the constraint
 $\Phi_{D+4}=x_o-\kappa\tau$ transforms into (\ref{C3}) after a
 canonical transformation from the variables $x^\mu,p_\mu$ to
 variables $x^{\prime\mu},p^\prime_\mu$, defined by the relations
 \begin{equation}
 \label{CTr}
 x^\prime_o=x_o-\kappa\tau,\quad x^{\prime i}=x^i,
 \quad p^\prime_\mu=p_\mu
 \end{equation}
Note also,  that we  omitted in  the complete  set of constraints
((\ref{C1})-(\ref{C5})) the pair  $\chi~\approx~0$,
$\pi_\chi~\approx~0$ ($\pi_\chi$  is  the  momentum,  canonically
conjugate  to   the variable $\chi$),  since the  Dirac brackets
for this  subset of constraints are equal to the Poisson brackets
for the variables of the theory.

To end this section, we will briefly comment on the
Bargmann-Michel-Telegdi equation for Pauli-Lubanski vector
$W_\mu$, which in $D=2n$ dimensions is defined as \cite{GG4}
\begin{eqnarray}
\label{PLVector}
&&W_\mu= \frac{(-i)^{\frac{D-2}{2}}}{(D-2)!}
\varepsilon_{\mu\nu\lambda_2\lambda_3\ldots \lambda_{D-1}}
{\cal P}^\nu \xi^{\lambda_2}\xi^{\lambda_3}\ldots \xi^{\lambda_{D-1}},\\
&&\hspace{3cm}{\cal P}^\nu=p^\nu-gA^{\nu,a}I^a.\nonumber
\end{eqnarray}
Going through the same steps as in \cite{GG5} we obtain the
desired equation
\begin{equation}
\label{EBMT}
\dot W_\mu=g\frac{M}{m}G^a_{\mu\nu}I^aW^\nu+
2Ru_\mu G^a_{\nu\lambda}I^aW^\nu
u_\lambda+O(\xi^{(D)}).
\end{equation}
where $u_\mu$ is the $D$-velocity: $u_\mu=\dot
x_\mu={\{x_\mu,H\}}^*$, $H=H_{\rm ext}\vert
_{\chi=0,e=1/m}$.

\section{Foldy-Wouthuysen Transformation and the ca\-no\-nical
variables}
\indent

As it was  mentioned in \cite{GG4},  it is convenient  to
quantize the theory not in terms of the initial variables, but in
terms of canonical variables, because the Dirac brackets of the
independent original  variables in the presence of the external
fields are  very   complicated  and the operator realization  of
the  theory in terms of the origional variables seems improbable.
The  canonical  variables  can be  found either by
diagonalization  of  the  Dirac  brackets  , as  it was done in
\cite{GG3},  or  by  using  pseudoclassical canonical Foldy   -
Wouthuysen transformation, as it was done in \cite{GG5,GG4}.  We
take   a   generator   of     the infinitesimal   canonical
transformations \cite{BCLFW} in the form

\begin{equation}
\label{Generator}
S=-2i\left({\cal P}_j \xi_j\right)\xi_{D+1}\theta(\gamma),
\end{equation}
Here  $\gamma=i{\{({\cal P}_i\xi_i),({\cal P}_j\xi_j)\}}^* =
 {\cal P}_i^2+igG^a_{ij}I^a\xi_i\xi_j,$ and
 $\theta(\gamma)$ satisfies the relation
$tg(2\theta\sqrt\gamma)=\frac{\sqrt\gamma}{\tilde{m}}$ ,
where $\tilde{m}=m-iRG^a_{ij}I^a\xi_i\xi_j$. The result of
the finite canonical  transformation  of   any   dynamical
quantity $A$   is given by the expression \cite{SM}
\begin{equation}
\label{CanonTrans}
\tilde{A}
=\widetilde {e^SA} = A+{\{A,S\}}^*+
\frac{1}{2!}{\{{\{A,S\}}^*,S\}}^*+ \ldots ,
\end{equation}
Repeating now the reasoning of \cite{GG4} one can prove the
following relation
\begin{eqnarray}
\label{PrimedDB}
{\{\tilde{A},\tilde{B}\}}_{D(\Phi)}&=&
{\{A^\prime,B^\prime\}}_{D(\Phi)}=\nonumber\\
&&={\{A,B\}}^*+
i{\{{\{A,B\}}^*,({\cal P}_i\xi_i)\}}^*
({\cal P}_j\xi_j)\frac{(b\kappa+a)}
 {\tilde \beta(\omega +\tilde m)},
\end{eqnarray}
where
\begin{equation}
\label{PrimedVar}
A^\prime \equiv \tilde A \vert_{\Phi = 0}=
A+i{\{A,({\cal P}_i\xi_i)\}}^*
({\cal P}_j\xi_j)\frac{(b\kappa+a)}
 {\tilde \beta(\omega +\tilde  m)},
\end{equation}
 $\tilde{\beta}=a\tilde{m}-b\kappa\omega$, $\omega = \sqrt{{{\cal
P}_i}^2+{\tilde m}^2+igG^a_{ij}I^a\xi_i\xi_j}$, and
$\{,\}_{D(\Phi)}$ denotes Dirac bracket for the complete set of
constraints (\ref{C1}-\ref{C5}).  The first equality in
(\ref{PrimedDB}) is the reflection of the property of the Dirac
brackets, according to which the  Dirac brackets of   the
constraints with any dynamical variable vanish.  The relations
 (\ref{PrimedDB}), (\ref{PrimedVar}) are crucial for finding the
canonical variables.  If now  we take  for $A,B$ the variables
$x_i,{\cal P}_j ,\xi_k, \theta_m$, then on account of
 (\ref{PrimedVar}) we find
\begin{equation}
\label{PIV1}
 x^\prime_i=x_i-i\xi_i ({\cal P}_j\xi_j)
 \frac{(b\kappa + a)}
 {\tilde{\beta} (\omega + \tilde m)} \equiv q_i,
\end{equation}
\begin{equation}
 \label{PIV2}
 {\cal P}_i^\prime={\cal P}_i +
igG_{ij}^aI^a\xi_j ({\cal P}_k\xi_k)\frac{(b\kappa+a)}
{\tilde \beta (\omega + \tilde m)}\equiv\pi_i,
\end{equation}
 \begin{equation}
\label{PIV3}
\xi_i^\prime = \xi_i + {\cal P}_i({\cal P}_j\xi_j)
\frac{(b\kappa+a)}
 {\tilde \beta(\omega+\tilde m)} \equiv \psi_i,
\end{equation}
\begin{equation}
\label{PIV4}
\theta_m^\prime=\theta^\prime-g(A_i^a\xi_i)T^a_{ml}\theta_l
({\cal P}_j\xi_j)\frac{(b\kappa+a)}
 {\tilde \beta(\omega+\tilde m)} \equiv\varphi_m.
\end{equation}
Applying now the formula (\ref{PrimedDB}) to variables $q_i,
\pi_j, \psi_i, \varphi_m $ we find the relations
\begin{equation}
\label{CV1}
{\{q_i,\pi_j\}}_{D(\Phi)}=
{\{x_i,{\cal P}_j\}}^*=-\delta_{ij},
\end{equation}
\begin{equation}
\label{CV2}
{\{\pi_i,\pi_j\}}_{D(\Phi)}
=gG_{ij}^a(x)I^a+ig\left(\nabla_k^{ab}G_{ij}^b(x)\right)I^a\xi_k
({\cal P}_m\xi_m)\frac{(b\kappa+a)}
{\tilde \beta(\omega+\tilde m)}=gG_{ij}^a(q)I^a_\varphi,
\end{equation}
\begin{equation}
\label{CV3}
 {\{\psi_i,\psi_j\}}_{D(\Phi)}=
{\{\xi_i,\xi_j\}}^*=-i\delta_{ij},
\end{equation}
\begin{equation}
\label{CV4}
\{\varphi_m,\varphi_n\}_{D(\Phi)}=
\{\theta_m,\theta_n\}^*=-i\delta_{mn},
\end{equation}
(all other brackets vanish).From these relations it is clear, that
the variables $q_i, \pi_j , \psi_k, \varphi_m $, which are just
the variables $\tilde x_i, \tilde {\cal P}_j, \tilde \xi_k,$
$\tilde{\theta}_m$ on the constraint surface, are canonical variables (they
  have canonical Dirac brackets).

For the variable $ I^{\prime a}\equiv I^a_\varphi=
\frac{1}{2}\varphi T^a\varphi$ one can find from (\ref{PrimedVar})
taking into account (\ref{IGC})
\begin{equation}
\label{Gen}
 I^{\prime a}=I^a-igf^{abc}(A_i^b\xi_i)I^c({\cal P}_j\xi_j)\frac{(b\kappa+a)}
 {\tilde \beta
 (\omega+\tilde m)},
\end{equation}
then from (\ref{PrimedDB}), with the account of (\ref{IGC})
and  (\ref{Gen}) one can deduce, that
\begin{equation}
\label{ICG}
{\{I^a_\varphi,I^b_\varphi\}}_{D(\Phi)}=f^{abc}I^c_\varphi.
\end{equation}

{}From formulae (\ref{PIV1}-\ref{PIV4}) one can find the expressions
 of  initial variables $x_i,{\cal P}_j~=~p_j-gA_j^aI^a , \xi_k, \theta_m$
 in terms of new variables $q_i, \pi_j=\Pi_j-gA_j^a(q)I^a , \psi_k,
  \varphi_m $ ($\Pi_i$ is the momentum, canonically conjugate to
  $q_i$):
\begin{equation}
\label{IV1}
 x_i  = q_i-i\psi_i(\pi_j\psi_j) \frac{(a\kappa + b)}
{\tilde \alpha(\Omega + m_\psi)},
\end{equation}
\begin{equation}
\label{IV2}
 {\cal P}_i=\pi_i + igG_{ij}^a(q) I^a_\varphi\psi_j
(\pi_k\psi_k)\frac{(a\kappa+b)}
{\tilde \alpha(\Omega + m_\psi)},
\end{equation}
\begin{equation}
\label{IV3}
\xi_i = \psi_i + \pi_i (\pi_j\psi_j)\frac{(a\kappa+b)}
{\tilde \alpha(\Omega+m_\psi)},
\end{equation}
\begin{equation}
\label{IV4}
\theta_m=\varphi_m-gT^a_{ml}\varphi_l(A_i^a(q)\psi_i)
(\pi_j\psi_j)\frac{(a\kappa+b)}
{\tilde \alpha(\Omega+m_\psi)},
\end{equation}
where  $\tilde \alpha=-a\kappa\Omega+bm_\psi,
  m_\psi= m-iRG_{ij}^a(q)I^a_\varphi\psi_i\psi_j,
\Omega=\sqrt{\pi_i^2+{m_\psi}^2+
igG_{ij}^a(q)I^a_\varphi\psi_i\psi_j}$.
The formula inverse to (\ref{Gen}) has the form
\begin{equation}
\label{IGen}
 I^a=I^a_\varphi-igf^{abc}(A_i^b(q)\psi_i)I^c_\varphi
 (\pi_j\psi_j)\frac{(a\kappa+b)}{\tilde \alpha(\Omega+m_\psi)},
 \end{equation}

 \section{Quantization}
 \indent

 The transition to the quantum theory is realized by the
replacement of the generalized coordinates and momenta by
operators, for which the commutators are defined by the rule
$[,]=i\hbar{\{,\}}_D$. Introducing operators $\hat{q}_i,
\hat{\Pi}_j , \hat{\psi}_k, \hat{\varphi}_m $ corresponding to
canonical variables $q_i, \Pi_j, \psi_k, \varphi_m $ we have

  \begin{equation}
  \label{ComRel}
[\hat q^i,\hat \Pi_j]_{-}=\hbar\delta^i_j,\quad
[\hat \psi^i,\hat \psi^j]_{+}
=\hbar\delta^{ij},
[\hat{\varphi}_m,\hat{\varphi}_n]_{+}=\hbar\delta_{mn}
\end{equation}
(all other brackets vanish).
The last two of these relations  generate  a  Clifford   algebras  in
$D-1$ and $N$ dimensional spaces correspondingly. The
finite dimensional matrix representations of these algebra are
given by the $2^{(D-2)/2}\bigotimes2^{(D-2)/2}$ matrices
$\sigma_i, \, i=1,\ldots,D-1$, and by
$2^{[N/2]}\bigotimes2^{[N/2]}$ matrices $\Sigma_m,\,m=1,\ldots,N$.
\begin{equation}
\label{SO}
\hat{\psi}_i=\left(\frac{\hbar}{2}\right)^{\frac{1}{2}}\sigma_i,
\quad\quad
\hat{\varphi}_m=\left(\frac{\hbar}{2}\right)^{\frac{1}{2}}\Sigma_m,
\end{equation}
 The isospin variable operator $I^a_\varphi$ is then given by
 \begin{equation}
\label{IO}
\hat{I}^a_\varphi=\frac{\hbar}{4}\Sigma_m T^a_{mn}\Sigma_n
\end{equation}
and belongs to the spinor representation of the algebra
$L_{SO(N)}$. Now, since , as it was mentioned above, the group $G$
 is a subgroup of $SO(N)$,  to find the isospin content of the
 quantum states one must decompose this spinor representation
 over the irreducible representations of $G$ \cite{BSSW}.

For the sake of completeness we will give the expression for the
quantum Hamiltonian of the theory
\begin{eqnarray}
\label{QH}
&&\hat{H}_{\rm
phys}\stackrel{\rightarrow}{\leftarrow}{\Omega}- g\kappa
A^a_0 I^a -ig \kappa\frac{G_{0k}^a I^a \psi_k (\pi_j
\psi_j)}{\Omega(\Omega+\tilde{m}_{\psi})}+\nonumber\\
&& +\frac{2iR\kappa}{\Omega}\left(G_{0k}^a I^a \psi_k+
\frac{\kappa G_{ik}^a I^a\pi_i\psi_k}{\Omega+\tilde{m_\psi}}\right)
(\pi_j\psi_j).
\end{eqnarray}
where $\stackrel{\rightarrow}{\leftarrow}$ mens the Weyl
correspondence between the symbols and their operators.
\vspace{5mm}
\noindent{\em Acknowledgment}
 This research was partially supported  by the grant 211-5291 YPI
 of the German Bundesministerium f\"ur Forschung und Technologie.
\newpage


\end{document}